\documentclass[epj,twocolumn]{webofc}
\usepackage[varg]{txfonts}   
\usepackage[final]{microtype}
\usepackage{graphicx}

\newcommand{\logg}{\ensuremath{\log g}\xspace}

\newcommand{\teff}{\ensuremath{T_{\textup{eff}}}\xspace}
\newcommand{\mlt}{\ensuremath{\alpha_{\textup{MLT}}}\xspace}
\newcommand{\mltg}{\ensuremath{\alpha_{\textup{MLT, grid}}}\xspace}
\newcommand{\radgrad}{\ensuremath{\nabla_{\textup{rad}}}\xspace}
\newcommand{\radgradt}{\ensuremath{\tilde{\nabla}_{\textup{rad}}}\xspace}
\newcommand{\dq}{\ensuremath{q\;\!'\,\!(\tau)}\xspace}
\newcommand{\dqq}{\ensuremath{q'(\tau)}\xspace}
\newcommand{\qt}{\ensuremath{q(\tau)}\xspace}
\newcommand{\taueff}{\ensuremath{\tau_{\textup{eff}}}\xspace}
\newcommand{\gar}{GARSTEC\xspace}

\wocname{epj}
\woctitle{Seismology of the Sun and the Distant Stars 2016}
\begin{document}
\title{Improving 1D Stellar Models with 3D Atmospheres}
%

\author{\firstname{Jakob Rørsted} \lastname{Mosumgaard}\inst{1}\fnsep\thanks{\email{jakob@phys.au.dk}} \and
        \firstname{V\' ictor} \lastname{Silva Aguirre}\inst{1} \and
        \firstname{Achim} \lastname{Weiss}\inst{2} \and
        \firstname{Jørgen} \lastname{Christensen-Dalsgaard}\inst{1} \and
        \firstname{Regner} \lastname{Trampedach}\inst{3,1}
}

\institute{%
  Stellar Astrophysics Centre (SAC),
  Department of Physics and Astronomy, Aarhus University, Ny
  Munkegade 120, DK-8000 Aarhus C, Denmark
  \and
  Max-Planck-Institut für Astrophysik,
  Karl-Schwarzschild-Str. 1, D-85748 Garching, Germany
  \and
  Space Science Institute, 4750 Walnut Street, Suite
  205, Boulder, CO 80301 USA
}

\abstract{%
  Stellar evolution codes play a major role in present-day
  astrophysics, yet they share common issues. In this work we seek to
  remedy some of those by the use of results from realistic and highly
  detailed 3D hydrodynamical simulations of stellar atmospheres. We
  have implemented a new temperature stratification extracted directly
  from the 3D simulations into the Garching Stellar Evolution Code to
  replace the simplified atmosphere normally used. Secondly, we have
  implemented the use of a variable mixing-length parameter, which
  changes as a function of the stellar surface gravity and temperature
  -- also derived from the 3D simulations. Furthermore, to make our
  models consistent, we have calculated new opacity tables to match
  the atmospheric simulations. Here, we present the modified code and
  initial results on stellar evolution using it.%
}
\maketitle
%
\section{Introduction}
\label{sec:intro}
Understanding stellar structure and evolution is one of the key
ingredients in astrophysics. Perhaps the primary tool for doing so is
evolutionary calculations of stellar structure, i.e. one-dimensional
numerical models. These have been developed and tested through
decades; as a result they are highly optimised and very
efficient. However, in several aspects they are also highly simplified
and can be improved. This work focuses on two of these: the use of a
simple temperature stratification as part of an idealised atmosphere
for the model and the treatment of stellar convection.

The artificial atmosphere is needed to supply the outer boundary
conditions for stellar evolution codes in order to solve the
differential equations of stellar structure (see e.g. \cite{kww}). The
boundary conditions are obtained by integrating an analytical
expression for the temperature as a function of optical depth -- a
so-called $T(\tau)$ relation. A general issue is that these relations
are much too simple to capture the true stratification of stellar
atmospheres.

In 1D models, convection is typically treated using a parametric
description. The most common is the mixing-length theory (MLT) as
introduced for stellar models by \cite{Vitense1953,Bohm-Vitense1958}
or some variant thereof. These parametric theories rely on rather
crude assumptions, which are certainly not fulfilled in stars (see
e.g. \cite{Trampedach2010}). Furthermore, a free parameter (the
mixing-length parameter, \mlt) is ``calibrated'' against the Sun by
some observational constraint, usually the solar radius. The solar
value is then assumed to be valid for all stars across the HR diagram.

To improve the models, highly sophisticated three-dimensional
simulations of stellar atmospheres (see the next section) can be used
in different ways. One approach is to patch the averaged 3D structure
on top of a 1D model, as described by \cite{Rosenthal1999}. This of
course requires the 1D model to match the 3D simulation perfectly.

In this work, we employ a different approach, where condensed
information -- specifically targeted at the two issues mentioned above
-- is extracted from the simulations and interpolated to any point
between them. The advantage of this approach is that it can be used
in stellar structure models for full evolutionary calculations; the
patched models are static by nature. An implementation similar to this
work, however not fully consistent, was attempted by
\cite{Salaris2015}.

In section~\ref{sec:sim} we review the 3D simulations used in this
work and how useful information is extracted from them. Our
implementation is described in detail in section~\ref{sec:implement},
and finally we present a few initial results in
section~\ref{sec:results}.

%
\section{Simulations of Stellar Atmospheres}
\label{sec:sim}
Stellar atmospheres are very complex regions, where hydrodynamics and
radiative transfer interact. For a better understanding of this
important part of stars several authors have employed advanced 3D
simulations, which account for both hydrodynamics and radiative
transfer. These simulations have been used extensively to study the
convection and granulation in the outer parts of our Sun
(e.g. \cite{Stein1989,Stein1998}).

\begin{figure}[htbp]
  \centering
  \includegraphics[width=\hsize]{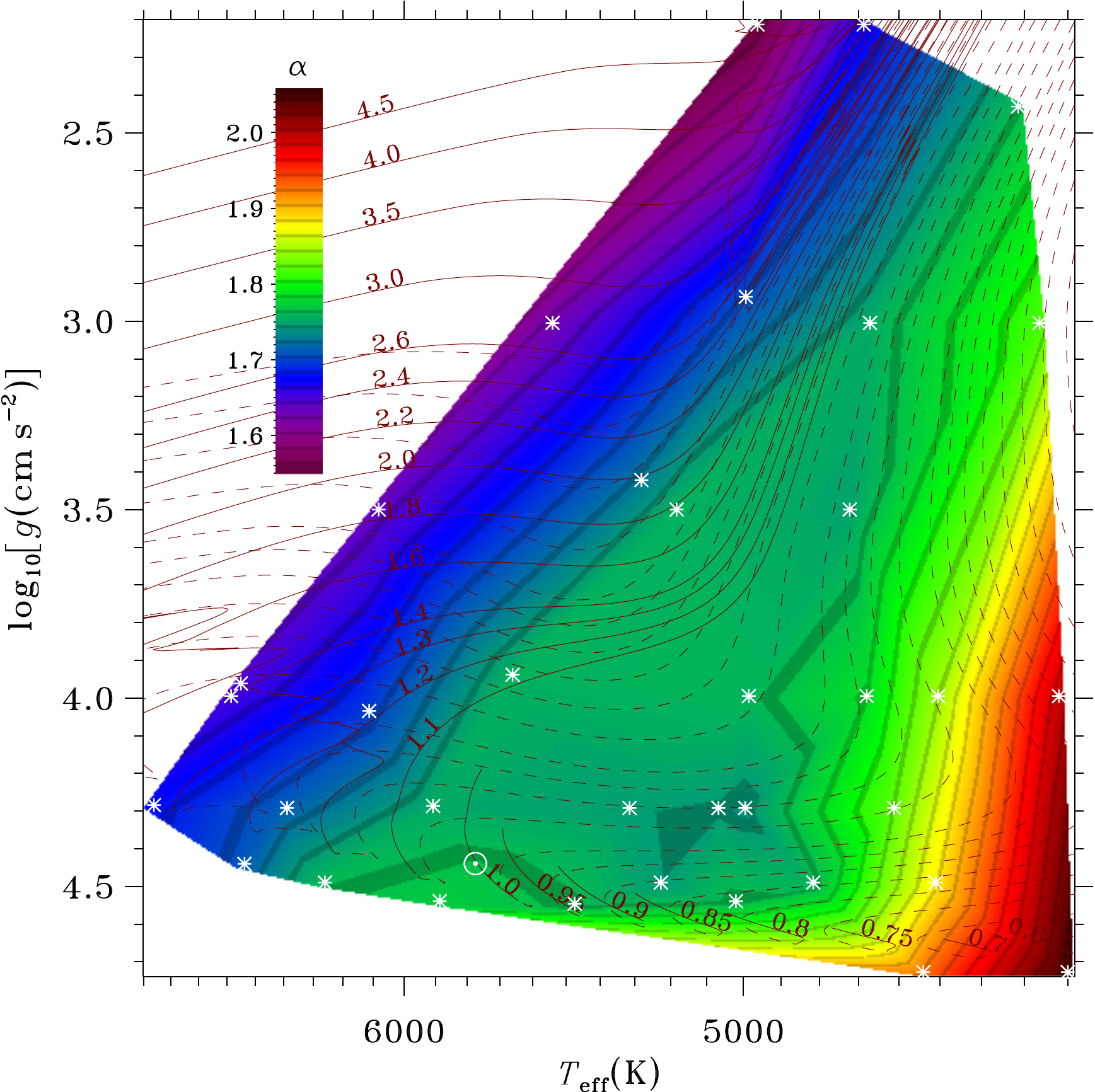}
  \caption{The grid of 3D simulations from \cite{Trampedach2013} at
    solar metallicity. The simulation parameters are marked by white
    asterisks and the Sun by $\odot$. The coloured background shows
    the behaviour of \mlt as a function of \teff and \logg. The red
    lines represents evolutionary tracks, but they are unimportant for
    the present work. The figure is reproduced from
    \cite{Trampedach2014} with permission.}
  \label{fig:grid}
\end{figure}

More recently, several groups have produced grids of 3D simulations to
investigate the behaviour of stars different from our Sun. In this
work, we will use the grid of simulations from Trampedach et
al. \cite{Trampedach2013}, which consists of 37 simulations at solar
metallicity -- the physical parameters (effective temperature, \teff,
and surface gravity, \logg) of the different simulations are shown in
Figure~\ref{fig:grid}, which is described in further details below.
%

\subsection{Condensed 3D Simulations}
\label{sec:sim_condens}
For the purpose of stellar evolutionary calculations, the full
(averaged) 3D structures cannot be used directly for several
reasons. First of all, they only sample discrete points in the HR
diagram, and it is not feasible to make the grids much denser since
the simulations are extremely computationally expensive.\footnote{One
  could in principle interpolate between the full simulations, but such
  a scheme has yet to be developed.} Secondly, due to differences in
physics (e.g. treatment of convection and turbulent pressure) they are
not directly compatible with the current 1D models.

For these reasons, Trampedach et al. \cite{Trampedach2014a} have
devised a way of distilling the 3D simulations to be easily usable in
stellar structure models. In short, the temperature stratification is
extracted from the simulations in the form of the (generalised) Hopf
function, $q$, as a function of Rosseland mean optical depth,
$\tau$. This is then used for reconstructing the photospheric
transition, from optically thick to optically thin, of the 3D
simulation in the 1D model.

Furthermore, it is possible to calibrate the mixing-length parameter,
\mlt, against the 3D simulations. This correctly connects the entropy
of the adiabatic layers deep in the star with the surface layers in
the 3D atmosphere. Besides, the need to assume a solar-calibrated
value for all stars is eliminated. This is done by
\cite{Trampedach2014} for all simulations in the grid -- the actual
procedure is very similar to the patching of 3D simulations to 1D
models mentioned earlier. The corresponding \mlt's are stored in the
table with \qt, and the interpolated behaviour as a function of \teff
and \logg is shown in Figure~\ref{fig:grid}. It should be noted that
the calibrated values are of course only strictly correct, if the
stellar evolution code uses the same MLT-formulation as the 1D
(envelope) code used for the calibration.

\section{Implementation}
\label{sec:implement}
\begin{figure}[htbp]
  \centering
  \includegraphics[width=\hsize]{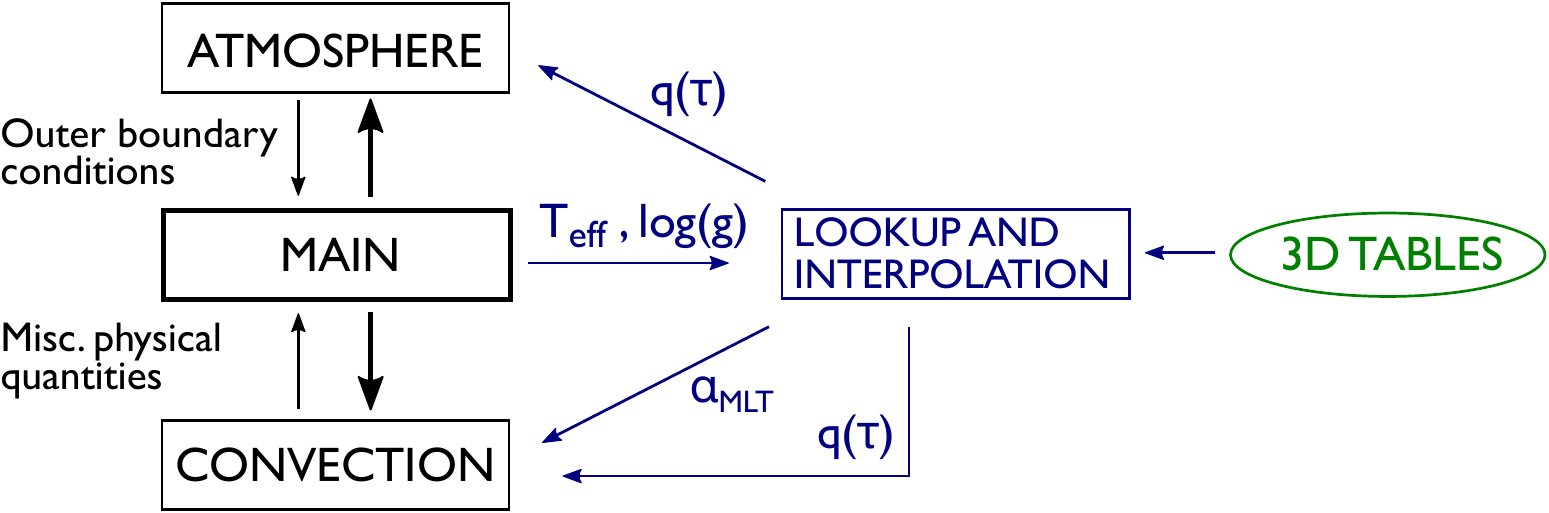}
  \caption{Schematic overview of our implementation marked in
    blue. The black represents modules in \gar, while the green
    circle marks the tables from
    \cite{Trampedach2014,Trampedach2014a}. Details are given in the
    text.}
  \label{fig:overview}
\end{figure}

We have implemented the results from
\cite{Trampedach2014,Trampedach2014a} (in the form of a table
containing \qt and \mlt) into the Garching Stellar Evolution Code
(\gar, see \cite{Weiss2008}). A simple sketch of the implementation
is shown in Figure~\ref{fig:overview}.

The general principle is that the \qt and \mlt corresponding to \teff
and \logg of the star are found by interpolation\footnote{We use the
  interpolation scheme supplied by \cite{Trampedach2014}, which is a
  linear interpolation over a Thiessen triangulation of the irregular
  grid of simulations.} in the grid and stored in the program. In each
iteration of the code, these values are updated to always match the
current stellar parameters.

The quantities are then supplied to the different parts of the code;
the atmospheric integration (providing the outer boundary conditions)
receives the \qt and the routine handling MLT-convection uses the new
\mlt. Moreover, for reasons explained below, the convection module
also needs access to \qt.
%

\subsection{Changes to a Stellar Model}
\label{sec:implement_model}
\begin{figure}[htbp]
  \centering
  \includegraphics[width=0.8\hsize]{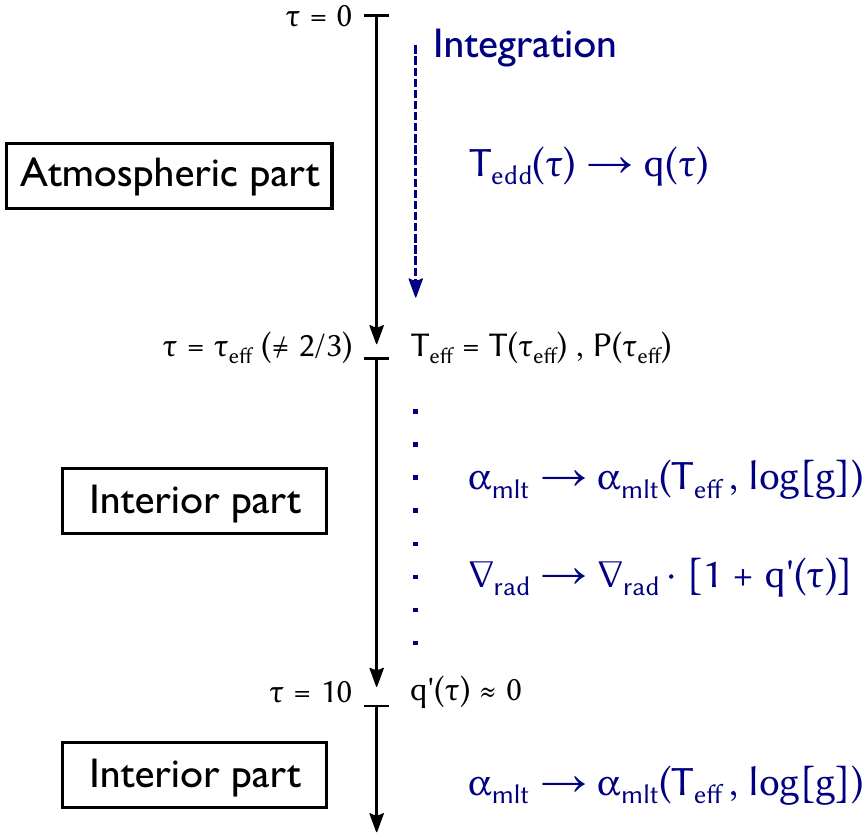}
  \caption{The changes to a stellar structure model as a result of our
    implementation. In the atmosphere, a different temperature
    stratification is used and the integration point is altered. In
    the outer, convective parts of the interior model, the radiative
    gradient is modified according to
    eq.~\eqref{eq:radgrad}. Everywhere in the interior of the star,
    the calibrated $\mlt(\teff,\logg)$ is used.}
  \label{fig:model}
\end{figure}

How a stellar structure model is modified by our implementation is
outlined in Figure~\ref{fig:model}.
%

In the atmosphere, the generalised Hopf function from the simulations,
\qt, is used to provide the stratification instead of the standard
analytical Eddington atmosphere, $T_{\textup{edd}}(\tau)$. Assuming
radiative equilibrium, the new temperature structure is found from \qt
using
\begin{equation}
  \label{eq:hopf}
  \frac{4}{3} \left( \frac{T}{\teff} \right)^{4} = \qt + \tau \; ,
\end{equation}
as explained by \cite[eq. 9]{Trampedach2014a}.

Another alteration is the bottom point of the atmosphere, i.e. where
the integration is stopped -- note that in stellar structure models
this point is usually defined to have $T = \teff$. The Eddington
atmosphere is anchored at $\tau = 2/3$, but that is not the case for
the relations extracted from 3D simulations. Thus before the actual
integration, we find the point \taueff in the interpolated temperature
structure where $T = \teff$.\footnote{The value is typically
  $\taueff \simeq 0.5$, and \teff is here the parameter of the
  (interpolated) 3D simulation.}

Moving on to the interior part of the model, the first change is the
use of the 3D-calibrated variable mixing-length parameter,
$\mlt(\teff, \logg)$, throughout the model. We are not using the \mltg
directly as provided from the tables, but rather using it
differentially as recommended by \cite{Trampedach2014a,Ludwig1999}. In
other words, we introduce a scaling factor
\begin{equation}
  \label{eq:mlt}
  \mlt(\teff, \logg) = \frac{\alpha_{\odot}}{\alpha_{\odot , \textup{ grid}}} \cdot
  \mltg(\teff, \logg) \; ,
\end{equation}
where $\alpha_{\odot}$ is what we obtain from a solar calibration in
\gar and $\alpha_{\odot , \textup{ grid}}$ is the value given for
the solar model in the grid.
This ensures that the solar model is calibrated to the correct radius
with this variable $\mlt$.

When using Eq.~\eqref{eq:hopf} for the temperature structure, the
temperature gradient for a stratification in radiative equilibrium
(i.e. no convective flux),
$\radgrad\equiv(\partial\ln T/\partial\ln p)_{\textup{rad}}$, needs
to be corrected from its optically deep value of
\radgradt, as
\begin{equation}
  \label{eq:radgrad}
  \radgrad = \radgradt \cdot [1 + \dq] \; ,
\end{equation}
according to \cite[eq. 35]{Trampedach2014a}. Here \dqq is the
derivative of the Hopf function with respect to $\tau$ and \radgradt
is the usual expression for the radiative gradient, based on the
diffusion approximation.  This modification is applied before entering
the MLT routine of \gar; thus the resulting temperature gradient
$\nabla$ is properly corrected.\footnote{The modified \radgrad is of
  course also used when checking for convection with the Schwarzschild
  criterion.} We apply this correction factor until $\tau = 10$, since
\dqq at larger depth is always below $10^{-4}$ (and typically below
$10^{-5}$), where the adiabatic layers are reached.

\subsection{Important Remarks}
\label{sec:implement_notes}
At this point it important to stress that it is crucial for the
stellar evolution code to use the same microphysics as the 3D
simulations for consistency.

To match the simulations, we use the MHD equation of state (EOS)
\cite{Hummer1988,Mihalas1988,Dappen1988}. To cover the necessary range
of temperatures, we complement it with OPAL EOS \cite{Rogers1996} at
higher temperatures.

For the 3D simulations, \cite{Trampedach2014a} calculated their own
atmospheric (i.e. low-temperature) opacities, which in the envelope
model used for the matching was merged with interior opacities from
the Opacity Project (OP, see \cite{Badnell2005}). We have calculated
similar opacity tables for the specific solar mixture used in the
simulations (see \cite{Trampedach2013}) using the atmospheric
opacities provided by the author and the available OP data.

\section{Initial Results}
\label{sec:results}
The natural test of our implementation is to calculate evolutionary
tracks and compare to a \emph{standard} evolution. In this context
\emph{standard} means: using Eddington atmosphere and a constant
solar-calibrated \mlt.

We did a solar calibration (using the opacities explained above) and
obtained $\alpha_{\odot} = 1.742$ for the traditional Eddington
atmosphere and $\alpha_{\odot} = 1.870$ for our new atmosphere. In the
latter case, eq.~\eqref{eq:mlt} is utilised to obtain the scaled \mlt
used in the calculation.\footnote{The solar value of the grid from
  \cite{Trampedach2014a} is $\alpha_{\odot , \textup{ grid}} = 1.767$,
  which gives a scaling factor of $1.058$ to be applied to the
  interpolated $\mlt(\teff, \logg)$.}

\begin{figure}[htbp]
  \centering
  \includegraphics[width=\hsize]{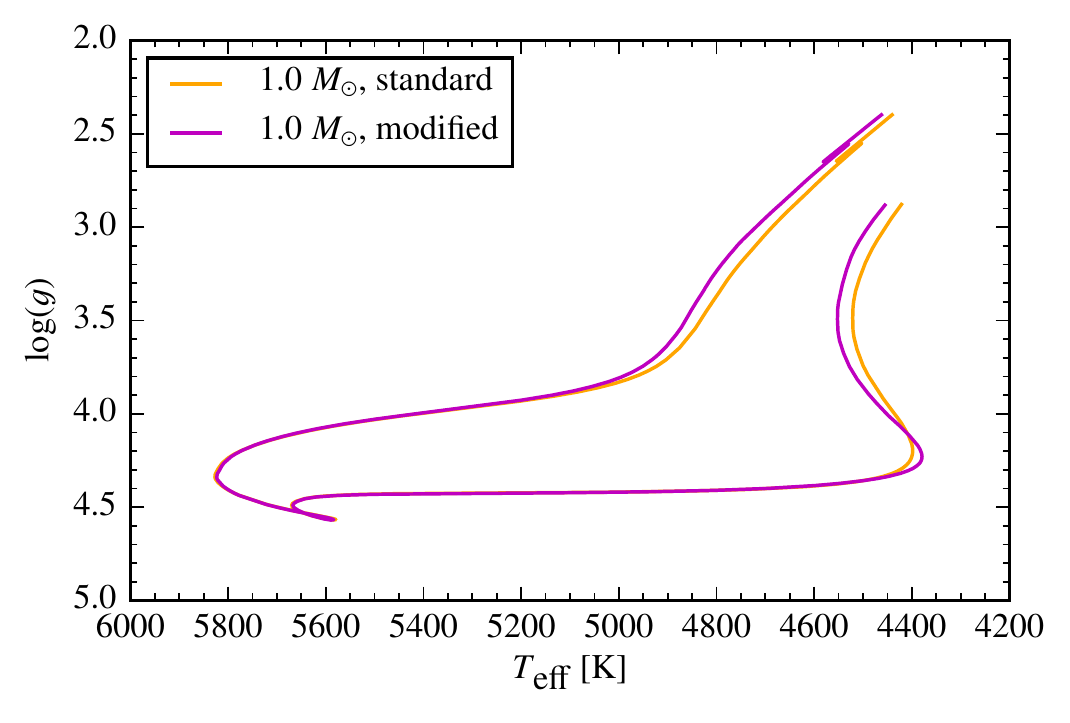}
  \caption{Evolutionary track of a $1.0 M_{\odot}$ star from the pre
    main sequence to $\logg = 2.4$. The \emph{standard} track uses a
    Eddington atmosphere and solar calibrated \mlt. The
    \emph{modified} track uses our implementation of \qt and
    $\mlt(\teff,\logg)$.}
  \label{fig:hrd100}
\end{figure}

\begin{figure}[htbp]
  \centering
  \includegraphics[width=\hsize]{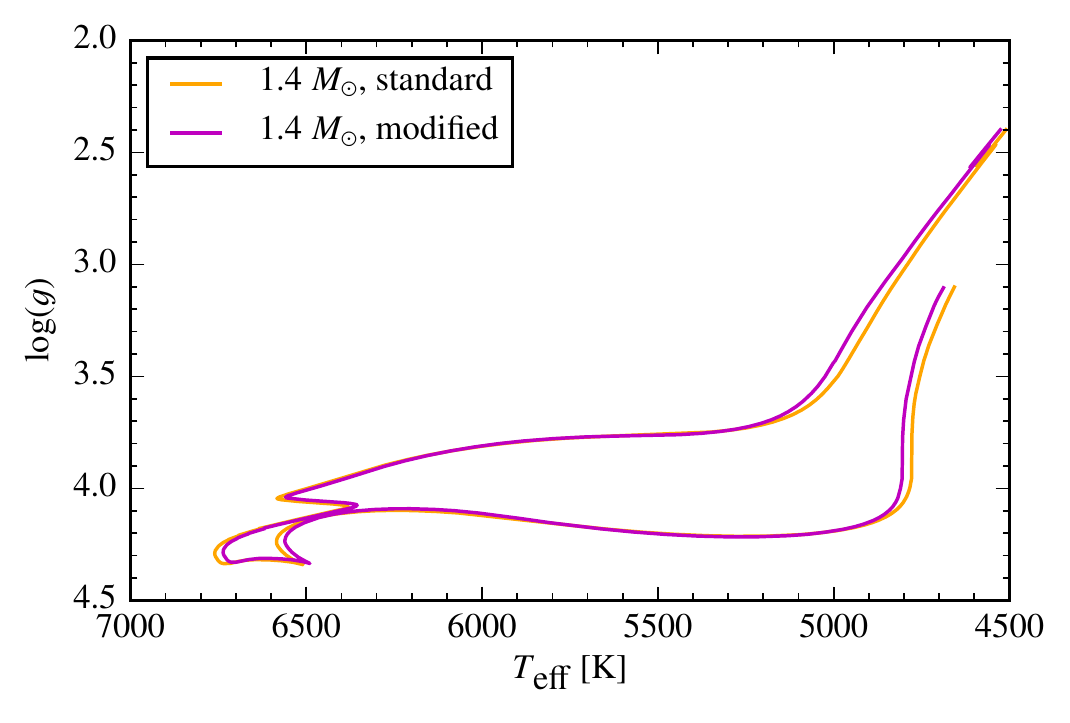}
  \caption{As Figure~\ref{fig:hrd100}, but for a $1.4 M_{\odot}$ star.}
  \label{fig:hrd140}
\end{figure}

The results for a $1.0 M_{\odot}$ and a $1.4 M_{\odot}$ star are shown
in Figures~\ref{fig:hrd100} and \ref{fig:hrd140}, respectively. It can
be seen that for the first track the temperature differences are
minuscule except at the RGB. For the heavier star, the differences are
larger, but the tracks are actually closer to each other higher up the
RGB.
%

Regarding the evolutionary pace of the models, it is also quite
similar. At the end of the main sequence (i.e. where central hydrogen
is exhausted) the age difference is less than $0.01$ Gyr between the
tracks -- similar for the end of the calculation at $\logg =
2.4$. However, using a more observational ``definition'' of the
turn-off -- the point of the track where \teff reaches maximum --
there is a difference in age of $\sim 0.5$ Gyr for the $1.0 M_{\odot}$
model. This is an interesting aspect of the new atmosphere, which
we will investigate further.

We plan to next look for differences in interior structure (if any)
between models with our implementation and the standard
calculation. It will be very interesting to use asteroseismology (see
e.g.  \cite{pulsbook}) and compare the oscillation frequencies between
modified and unmodified models. We will follow this line of
investigation in a coming paper.

\section{Conclusion}
\label{sec:conclusion}

We have successfully implemented $T(\tau)$ relations extracted from 3D
simulations and the associated modifications to the thermal gradients
in the Garching Stellar Evolution Code. Furthermore, we have also
modified the code to use a variable mixing-length parameter,
$\mlt(\teff,\logg)$, obtained from calibration against the 3D
simulations. For consistency we have calculated new opacity tables to
match those used in the simulations, and we use the same equation of
state.

In a future work we will investigate the effect of our modifications
on the interior structure of the models and more specifically the
oscillation frequencies. We will also investigate the impact on the
derived age when comparing models to observations.

\section*{Acknowledgements}

Funding for the Stellar Astrophysics Centre (SAC) is provided by The
Danish National Research Foundation (Grant agreement no.: DNRF106).

\phantom{Hest}

\bibliography{JakobMosumgaard.bib}

\end{document}